\documentclass[aps,prb,reprint,amsmath,amssymb,superscriptaddress,twocolumn,nofootinbib,longbibliography]{revtex4-2}
\usepackage{graphicx}
\usepackage{dcolumn}
\usepackage{bm}
\usepackage{physics}
\usepackage{graphicx}
\usepackage{subfigure}
\usepackage{braket}
\usepackage{amsmath}
\usepackage{commath}
\usepackage{mathtools}
\usepackage{float}
\usepackage{placeins} 
\usepackage{chemformula}
\usepackage[normalem]{ulem}
\usepackage{wasysym}
\usepackage{txfonts}
\usepackage{tikz}
\usepackage{orcidlink}
\usepackage{mathrsfs}
\usepackage{makecell}
\usepackage{array}
\usepackage{tabularray}

\begin{document}
	
	\preprint{APS/123-QED}
	
    \title{Nonlocal Transport in Cr-doped (Bi,Sb)$_2$Te$_3$: Absence of Nonchiral Edge States}
	
	\author{Valery Ortiz Jimenez}
	\email{Contact author: valery.ortizjimenez@nist.gov}
	\affiliation{Physical Measurement Laboratory, National Institute of Standards and Technology, Gaithersburg, MD 20899, USA.} 
        \author{Paul M. Haney}%
        \email{Contact author: paul.haney@nist.gov}
        \affiliation{Physical Measurement Laboratory, National Institute of Standards and Technology, Gaithersburg, MD 20899, USA.}	
        \author{Farzad Mahfouzi}%
        \affiliation{Physical Measurement Laboratory, National Institute of Standards and Technology, Gaithersburg, MD 20899, USA.}	
        \author{Ngoc Thanh Mai Tran}%
        \affiliation{Associate, Physical Measurement Laboratory, National Institute of Standards and Technology, Gaithersburg, MD 20899, USA.}	
        \affiliation{Joint Quantum Institute, University Of Maryland, College Park, College Park, Maryland, 20742, USA}
        \author{Albert F. Rigosi}%
        \affiliation{Physical Measurement Laboratory, National Institute of Standards and Technology, Gaithersburg, MD 20899, USA.}	
	\author{Curt A. Richter}%
    \email{Contact author: curt.richter@nist.gov}
	\affiliation{Physical Measurement Laboratory, National Institute of Standards and Technology, Gaithersburg, MD 20899, USA.}	
	
	\begin{abstract}
    The quantum anomalous Hall effect shows great promise for realization of the ohm without the need for an external magnetic field. The most mature material platform is magnetically doped topological insulators. In these materials, precise quantization is limited to low temperatures, with the activation energy for dissipative transport typically in the range of $1~{\rm K}$. One potential source of dissipative transport is non-chiral edge states.  These states are expected to be present in sufficiently thick samples.  In this work, we perform extensive Hall and non-local resistance measurements in a Hall bar geometry at 2 K.  By comparing 15 independent transport measurements to different transport models, we find that the system behavior is well-described by a simple continuum Ohm's law model.  The addition of non-chiral edge states into the model does not significantly improve the fitting, and we conclude that there is not strong evidence for these states.  We discuss the implications of our results for the prospect of high-temperature quantized anomalous Hall effect in these materials.
    

	\end{abstract}
	
	\maketitle
	
	
	\section{\label{sec:sec1}Introduction}

The quantum anomalous Hall effect is a quintessential manifestation of topology in condensed matter systems \cite{chang2023colloquium}. It leads to a Hall resistance which is quantized at the von Klitzing constant $h/e^2$ (where $h$ is Planck's constant and $e$ is the elementary charge) in the absence of an applied magnetic field.  Efforts to utilize the quantum anomalous Hall effect as a resistance standard have made substantial progress, with relative combined standard uncertainty reaching values of $10^{-9}$ \cite{okazaki2022quantum,patel2024zero,rodenbach2022metrological}.  A key advantage of the quantum anomalous Hall effect over the quantum Hall effect is that it requires no external magnetic field.  This is particularly attractive because it enables integration with the voltage standard, which requires superconducting Josephson junction and is therefore incompatible with the large magnetic fields needed for the quantum Hall effect \cite{huang2025quantum}.  Indeed, the combination of the quantum anomalous Hall effect resistance and Josephson junction volt was recently utilized to realize the ampere \cite{rodenbach2025unified}.

The most well-developed material class to exhibit the quantum anomalous Hall effect are magnetically doped topological insulators.  These materials are inherently disordered and therefore exhibit rather low thermal activation energies for longitudinal transport, on the order of 1~${\rm K}$ to 2~${\rm K}$ \cite{fox2018part,okazaki2022quantum,pan2020probing,footnote}. For this reason, they exhibit metrologically relevant quantum anomalous Hall resistance values at low temperatures, on the order of tens of mK.  Pushing the operating temperature to $\approx$2~${\rm K}$ would markedly reduce the cryogenic overhead and enable a smaller cooling footprint \cite{callegaro2025quahmet}. 

\begin{figure}%
	{\includegraphics[scale=0.2,angle=0,trim={0.0cm 0.0cm 0.0cm 0.0cm},clip,width=.48\textwidth]{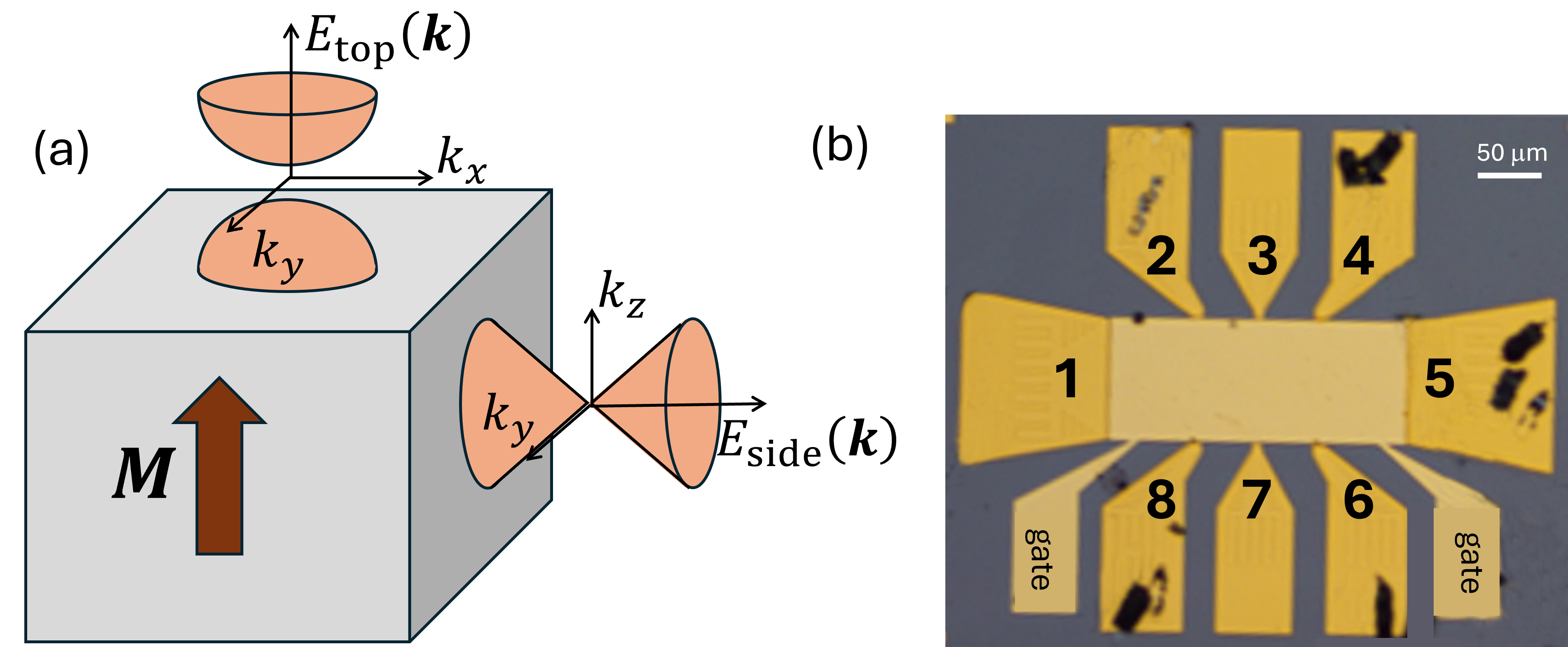}}%
	\caption{(a) Schematic of a 3-dimensional Chern insulator with magnetization along the $z$-direction.  Surfaces with normal along $z$ are gapped, while other surfaces remain gapless.  (b) Experimental sample with lead index labeling.}%
	\label{fig:fig1}%
\end{figure}

In addition to bulk disorder, a potential source of non-ideal behavior in these materials is the possible presence of non-chiral edge states \cite{wang2013anomalous}. The occurrence of these states can be anticipated from the underlying material properties, as we discuss next.  The ``recipe'' for these materials begins with a topological insulator, which, in 3-dimensions, possesses non-chiral surface states with a Dirac dispersion on all of its surfaces.  The material is then doped with magnetic atoms, and under appropriate conditions, a magnetically ordered state can emerge.  The Dirac cone on the surface whose normal direction is collinear with the magnetization is gapped, while other surfaces may remain gapless (see Fig 1(a)).  These ``side-wall'' surface states are non-chiral and can backscatter, introducing dissipation and reducing the Hall quantization.  The details of non-chiral side-wall states will be highly dependent on sample thickness.  For quantum anomalous Hall devices, the thickness is typically on the order of several quintuple layers (QL), so that the spectrum of side wall surface states is discretized from quantum confinement effects, or removed altogether due to hybridization of top/bottom surfaces.  In the latter case, the system enters a proper 2-dimensional regime.  For topological insulators such as Bi$_2$Se$_3$, this transition occurs at around 6 QL \cite{zhang2010crossover}.

Previous works have argued for the presence of non-chiral edge states in quantum Hall devices based on transport measurements \cite{kou2014scale,wang2020demonstration}.  These experiments considered thicker samples with thicknesses of 10 to 11 QL, facilitating the formation of non-chiral edge states on the sidewalls.  The device thickness for metrology applications are notably smaller, typically 5 to 6 QL.  However, previous work has also argued that thinner devices (4 QL-thick) can exhibit thermally activated non-chiral edge state transport \cite{chang2015zero}. This indicates the potentially important role of these states in the quest for high-temperature quantum anomalous Hall devices. Other experiments have attributed deviations from quantization solely to bulk transport \cite{rodenbach2021bulk,rosen2022measured}, and imaged the bulk current directly~\cite{ferguson2023direct}.  Experiments on samples with thicknesses of up to 106 nm showed no evidence for non-chiral states at low temperature ($\approx 25~{\rm mK}$) \cite{zhao20243d}. It's therefore important to clarify the relevance of these states to device behavior for relevant thicknesses, and at the elevated temperatures sought after for next-generation devices.


%

In this work, we perform transport measurements on a magnetically doped topological insulator to determine the role of non-chiral edge states on the transport.  Previous works often use gate voltage and temperature as the experimental knobs to discern possible transport mechanisms.  In this work, we use measurement configuration as our primary experimental parameter.  The motivation for this is that different gate voltages and temperatures can change the nature of the charge transport, which complicates the model fitting procedure: different measurement parameters can lead to different model parameters, or even different models. On the other hand, through the symmetry analysis presented here, we show that there are 17 independent resistance measurements available on our 8-terminal Hall bar.  We present an analysis of 15 independent measurements, and critically assess the fitting of our data to transport models with and without non-chiral edge states.  We find that a two-parameter continuum Ohm's law model, as described in Ref. \cite{rosen2022measured}, provides the best description of the data, and that there is therefore not strong evidence for non-chiral edge states in our sample.

\section{experiment}

The device presented in this work is a heterostructure of four QLs of Cr$_{0.12}$(Bi$_{0.26}$Sb$_{62}$)$_{2}$Te$_{3}$ grown between a top and bottom QL of Cr$_{0.24}$(Bi$_{0.22}$Sb$_{54}$)$_{2}$Te$_{3}$. The Cr-doped BST structure was grown in an ultra-high vacuum molecular beam epitaxy system, on a GaAs (111) B substrate annealed at $630 ^{\circ}$C. The device was fabricated via direct-write optical photolithography. The device geometry was defined via Ar ion milling and shaped into a Hall bar. Electrical contacts, composed of a $5~$nm Ti layer, and a $90~$nm Au layer, were deposited via electron beam evaporation. The top gate was formed by depositing a $1~$ nm Al layer which was subsequently oxidized, followed by a deposition of a $40~$nm AlO$_2$ layer via  atomic layer deposition.   

Direct current (dc) electrical measurements were taken in a $^4$He cryostat system, with a temperature range from 1.6 K to 400 K, and magnetic field capabilities up to 14 T. An 8.5 digit multimeter was used to measure the DC voltage drop between two leads along the path of an applied DC current (I = 50 nA) through the device. The current is directly measured by another 8.5 digit multimeter. The gate voltage of the device was set to -0.7 V, which was determined to locate the Fermi level near the center of the exchange gap. To improve accuracy, the reported resistance values are the average results of two current directions, measured over 50 power line cycles. While slightly noisier relative to widely used lockin measurements, the 8.5 digit dc measurements are more accurate. In addition, we observe that some conventional lockin measurements do not satisfy the symmetry constraints outlined in the next section.   We attribute this to alternating current (ac) effects, which are present even at very low frequencies; this is a topic for future investigation. It should be noted that specialized metrological equipment can be used to determine the resistance more accurately (see, for example, \cite{fox2018part}.) However, such high-accuracy measurements are not necessary for this investigation. \\

We present results in terms of the resistance between two voltage tabs of the device, which we define as:
\begin{eqnarray}
    R_{ij,k\ell} = \frac{I_{ij}}{V_k - V_\ell}
\end{eqnarray}
where $I_{ij}$ is the applied current between source and drain leads $i$ and $j$, and $V_k-V_\ell$ is the measured potential difference between leads $k$ and $\ell$.  Note that in the following sections, we use the notation $(i,j;k,\ell)$ to denote the measurement configuration for the resistance $R_{ij,k\ell}$.

Fig.~\ref{fig:fig2} shows the resistance values for 9 different measurement configurations as a function of applied magnetic field.  All measurements are taken at $T\approx2~{\rm K}$.  For the purposes of comparing different transport models, we focus on the high and low field saturated resistance values.  Panels (h) and (i) show  configurations corresponding to the conventional Hall and longitudinal resistance measurements, respectively, while the other configurations are 3 or 4-terminal non-local resistance measurements. 

\begin{widetext}
\begin{center}
\begin{figure}%
	{\includegraphics[scale=0.2,angle=0,trim={0.0cm 0.0cm 0.0cm 0.0cm},clip,width=.8\textwidth]{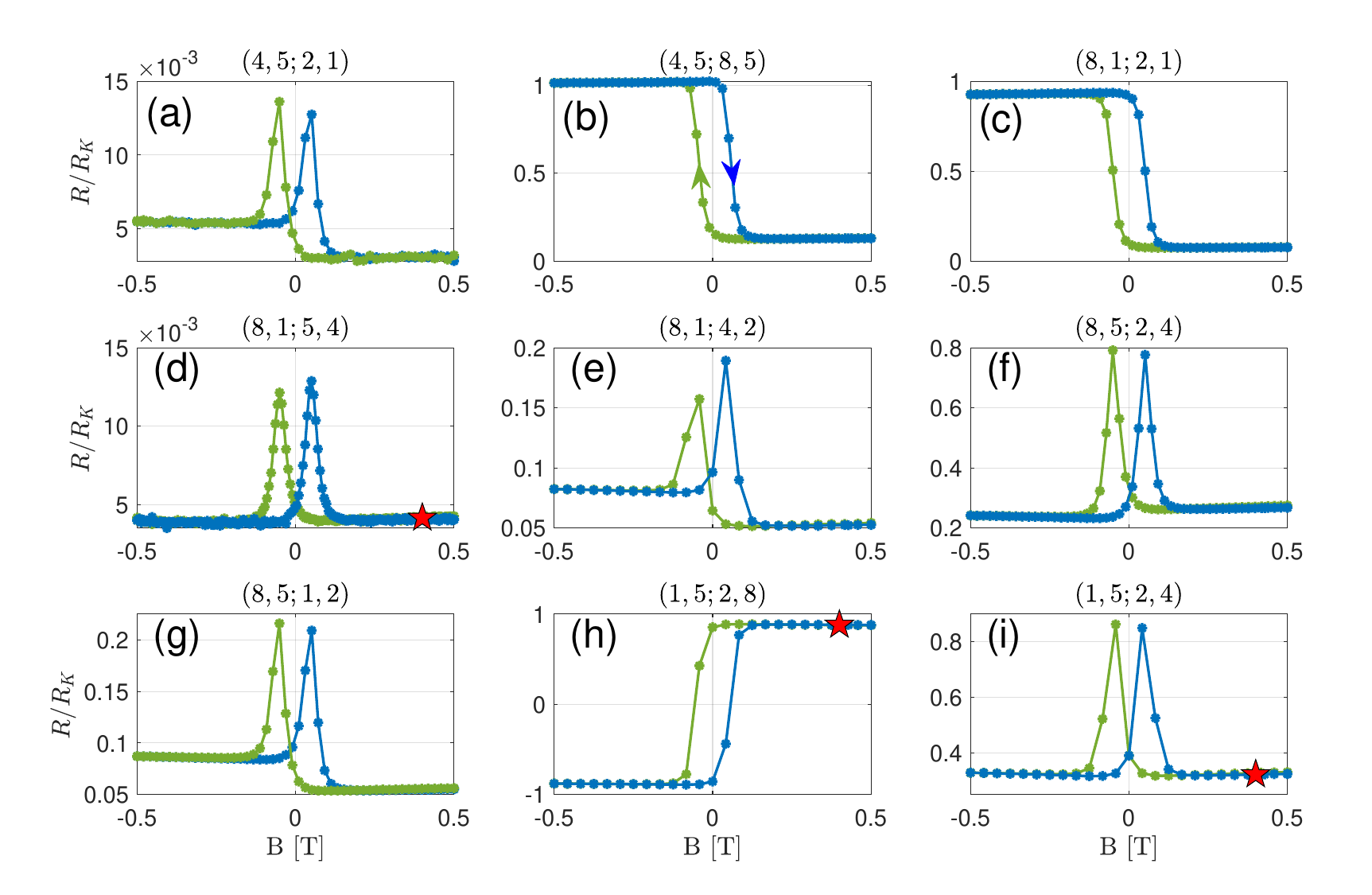}}%
	\caption{Experimental data of resistance for different measurement configurations.  Blue and green curves show upward and downward $B$-field sweeps, respectively, as indicated by the arrows in (b).  The red stars in (d), (h), and (i) indicate symmetry-equivalent measurement configurations at $+M$ and $-M$. The blue (green) solid lines correspond to upward (downward) sweeping of the magnetic field. The subplot titles $(i,j;k,\ell)$ indicate the current source/drain lead $i$ and $j$, and leads $k$ and $\ell$ across which voltage is measured.  See Fig.~\ref{fig:fig1}(b) for lead labeling. The $y$-axis is the resistance $R_{ij;k\ell}$ scaled by the von Klitzing constant $R_K$.}
	\label{fig:fig2}%
\end{figure}
\end{center}
\end{widetext}


\section{Models}

\subsection{Symmetry analysis}

In this section, we provide a symmetry analysis of the system, which provides rigorous constraints on the system response.  The primary experimental knob in this work is the applied current lead pair and the measured voltage lead pair.  In an 8-terminal device, there are nominally ${8\choose2}\times {8\choose 2}=784$ measurement choices.  However, by virtue of the system symmetry and the assumption of linear response, many of these measurements are symmetry-equivalent.  The symmetry analysis we present below shows that there are 17 symmetry-independent measurements available for this system.

We begin by specifying the notation to describe our measurements. As described in the previous section, the 4-tuple $(i,j;k,\ell)$ specifies the current lead-pair $i,j$ and voltage lead-pair $k,\ell$.
We vary the magnetic field to align the magnetization to both out-of-plane orientations.  We label the magnetization orientation with $\sigma$, which takes the value of $\pm 1$.  This is denoted with a subscript, so that a measurement is given by $(i,j;k,\ell)_\sigma$.

Next, we analyze the system symmetries and the resulting constraints on the measurement set.  The system shown in Fig.~\ref{fig:fig1} possesses the following symmetries: rotation by $180^\circ$ along the axis perpendicular to the plane ($C_2$), and the combined action of mirror operation about the $x$-axis or $y$-axis, denoted by $\mathcal{M}_x$ and $\mathcal{M}_y$, respectively, followed by the time-reversal operation $\mathcal{T}$: $\mathcal{M}_x \mathcal{T}$ and $\mathcal{M}_y \mathcal{T}$.  These three operations, together with the identity operation $I$ yield the generators of the system's symmetry group: $\{ I, C_2,\mathcal{M}_x \mathcal{T},\mathcal{M}_y \mathcal{T} \}$.  The mapping of the leads under the spatial operations are given by:
\begin{eqnarray}
C_2 (1,2,3,4,5,6,7,8) &=& (5,6,7,8,1,2,3,4)\\
\mathcal{M}_x  (1,2,3,4,5,6,7,8) &=& (1,8,7,6,5,4,3,2)\\
\mathcal{M}_y  (1,2,3,4,5,6,7,8) &=& (5,4,3,2,1,8,7,6)
\end{eqnarray}

Two measurements $(i,j;k,\ell)_\sigma$ and $(i',j';k',\ell')_{\sigma'}$ are symmetry-equivalent if there is a symmetry operation $U$ that maps the indices of one measurement to the indices of the other measurement: $i'=U(i),~ j'=U(j),~ k'=U(k),~ \ell'=U(\ell),~\sigma'=U(\sigma)$.  




An additional constraint on the system response is Onsager reciprocity. Onsager reciprocity states that the response is the same when switching the driving force index (in our case, the applied current lead pair) with the response index (in our case, the voltage lead pair), and performing time reversal.  In our notation, this reads: $(i,j;k,\ell)_\sigma=(k,\ell;i,j)_{-\sigma}$.  


Additional constraints are derived from linearity.  If the current is applied between two leads $i$ and $j$, and the voltage difference is measured between two sets of leads $k$ and $\ell$ and $\ell$ and $m$, then the sum of these two voltages would yield the measured voltage between $k$ and $m$.  Symbolically, this is represented as:  $(i,j;k,m)_\sigma =(i,j;k,\ell)_\sigma+(i,j;\ell,m)_\sigma$.  Linearity also implies that switching the order of current lead indices $i$ and $j$ (or, reversing the source and drain electrodes) or voltage lead indices $k$ and $\ell$ merely changes the sign of the measured voltage.  A switch of current indices and/or voltage indices is therefore not an independent measurement.  With these constraints, there are a total of 17 independent measurements for the system.  See Appendix~\ref{app:symmetry} for a derivation of this number.

For the purposes of illustration, we describe two symmetry-equivalent measurements.  Consider the non-local measurements of opposite corner pairs at $+M$ and $-M$: $(1,3;2,8)_+$ and $(1,3;2,8)_-$.  The first measurement can be transformed to the second as follows: apply Onsager reciprocity: $(1,3;2,8)_+ \xrightarrow[\text{}]{\text{Ons}} (2,8;1,3)_-$, and next apply a $C_2$ rotation: $(2,8;1,3)_- \xrightarrow[\text{}]{C_2} (1,3;2,8)_-$. The two measurements are therefore symmetry-equivalent: $(1,3;2,8)_+ \equiv (1,3;2,8)_-$.

 



\subsection{Landauer-Buttiker description of transport}
We describe electronic transport through the system with a Landauer-Buttiker picture of transport \cite{buttiker1986four}.  The current through lead $i$ is given by:
\begin{eqnarray}
I_i = \frac{e^2}{h}\sum_j T_{j,i} \mu_i - T_{i,j} \mu_j
\end{eqnarray}
where $T_{i,j}$ is the probability for a state injected in lead $j$ to be collected by lead $i$ ({\it i.e.}, the transmission probability from lead $j$ to $i$), and $\mu_i$ is the electrochemical potential of lead $i$.

A model is fully specified by its transmission matrix $T$, and a measurement is specified by the boundary conditions $I_i=-I_j=I_{\rm app}$, $I_k=0$ for $k\neq i,j$.  The electrochemical potentials satisfy:
\begin{eqnarray}
    I_i=  \frac{e^2}{h}G_{i,j} \mu_j \label{eq:IV}
\end{eqnarray}
where
\begin{eqnarray}
    G_{i,j} = \begin{cases}        
    T_{i,j}~~~~~~~~~~~~~~~~{\rm for~}i\neq j\\
    -\sum_{k} T_{ik}~~~~~~~{\rm for~}i=j
    \end{cases}
\end{eqnarray}
Eq.~\ref{eq:IV} is numerically solved by choosing a zero of energy; this amounts to setting the chemical potential of one lead to zero and removing this lead from the set of equations. The matrix is then non-singular and the solution is easily computed.

For a perfectly chiral system, $T_{i,i+1}=1$ and $T_{i,j}=0$ for $j \neq i+1$.  We adopt the convention that the leads are numbered in clockwise order (see Fig.~\ref{fig:fig1}(b)), so that for an $N$-terminal lead, index $N+1$ is mapped to $1$, and index $0$ is mapped to $N$.

\subsection{Non-chiral edge state model}

As discussed in the introduction, non-chiral edge states can be generically expected in a sufficiently thick magnetically doped topological insulator.  In this section, we review a model for non-chiral edge states and present their impact on the quantization of the anomalous Hall effect.  We find that these states can lead to rapid deterioration of the Hall quantization, particularly with respect to the levels relevant for metrology applications.  This is a primary motivation for carefully diagnosing the presence or absence of non-chiral edge states.

To model co-existing chiral and non-chiral edge states in a Landauer-Buttiker model, we let $T_{i,i+1} =1+k,~T_{i,i-1}=k$. (Previous models~\cite{wang2013anomalous} consider different transmission probabilities of the non-chiral edge state to the two adjacent leads, introducing two parameters $k_1$ and $k_2$.  Here we simplify the model and consider $k_1=k_2=k$.)  Ref. \cite{wang2013anomalous} provides an analytical expression for the voltage of a 6-terminal device in a conventional Hall measurement.  Here, we provide a general solution for an injected current lead pair $i,j$.  For $V_i=0,~V_j= 1$ and $i<j$, the potential at the remaining leads $k \neq i,j$ is given by:
\begin{eqnarray}
    V_k = \begin{cases}        
    
    \left(\dfrac{1 - r^{\left(k-j\right)}}{1-r^{\left(j-i\right)}}\right)~~~~~~~~~~~~~~~~{\rm for~}i<k<j\\
    \left(\dfrac{1-r^{-{\rm mod}\left(i-k,N\right)}}{1-r^{-{\rm mod}\left(i-j,N\right)}}\right)~~~~~~~~~{\rm otherwise}
    \end{cases} \label{eq:non-chiral}
\end{eqnarray}
where $r=k/(1+k)$.  

The impact of non-chiral edge states on the quantization of the anomalous Hall resistance can be found directly from Eq.~\ref{eq:non-chiral}.  As discussed in previous works, the number of leads entering the expression is not necessarily equal to the number of physical leads on the device.  ``Virtual'' Buttiker probes are also included in this model to describe phase-breaking scattering.  The total number of leads along an edge is therefore the sum of the number of physical leads and the number of phase-breaking scattering events when transiting from source to drain electrode along an edge. Let the total number of leads along both edges (and source/drain contacts) be $N$. For a conventional Hall measurement, the Hall resistance in the middle of the device is:
\begin{eqnarray}
    R_{xy} = \frac{h}{e^2}\left( \frac{2}{1+r^{N/4}}-1\right)
\end{eqnarray}
The above equation is derived when the total number of leads $N$ is divisible by 4.  However, the overall scaling should be applicable for any number of leads.  

\begin{figure}%
	{\includegraphics[scale=0.2,angle=0,trim={0.0cm 0.0cm 0.0cm 0.0cm},clip,width=.5\textwidth]{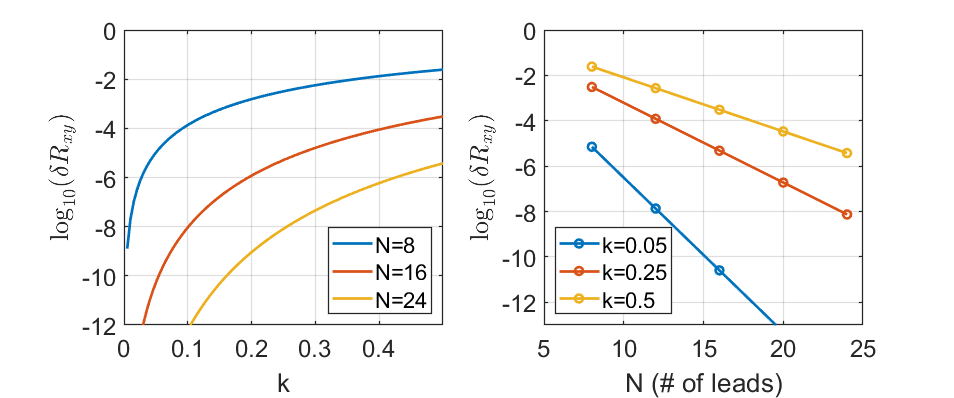}}%
	\caption{Deviation of the Hall resistance in the center of the device, $\delta R_{xy}$, from the von-Klitzing constant due to non-chiral edge states ($\delta R_{xy}=1-R_{xy}/R_K$). (a) shows the variation of $\delta R_{xy}$ with non-chiral edge scattering parameter $k$, for varying number of leads as shown in the legend. (b) shows the same versus number of leads, for various $k$ values as shown in the legend.}
	\label{fig:non-chiral}%
\end{figure}

The reduction in quantization is shown in Fig.~\ref{fig:non-chiral}, as a function of $k$ and number of leads. For a wide range of parameter values, the reduction is significant, particularly for metrology applications. The presence of non-chiral edge states is therefore an important consideration for device performance, particularly at higher temperatures. One notable feature of non-chiral edge states is that their impact on quantization is mitigated by adding more scattering (or increasing $N$).  The introduction of phase-breaking scattering suppresses the non-chiral contribution to transport, while the chiral edge modes remain topologically protected and thus unaffected.

\subsection{Continuum Ohm's law model}

\begin{figure} [h!]
	{\includegraphics[scale=0.2,angle=0,trim={0.0cm 0.0cm 0.0cm 0.0cm},clip,width=.45\textwidth]{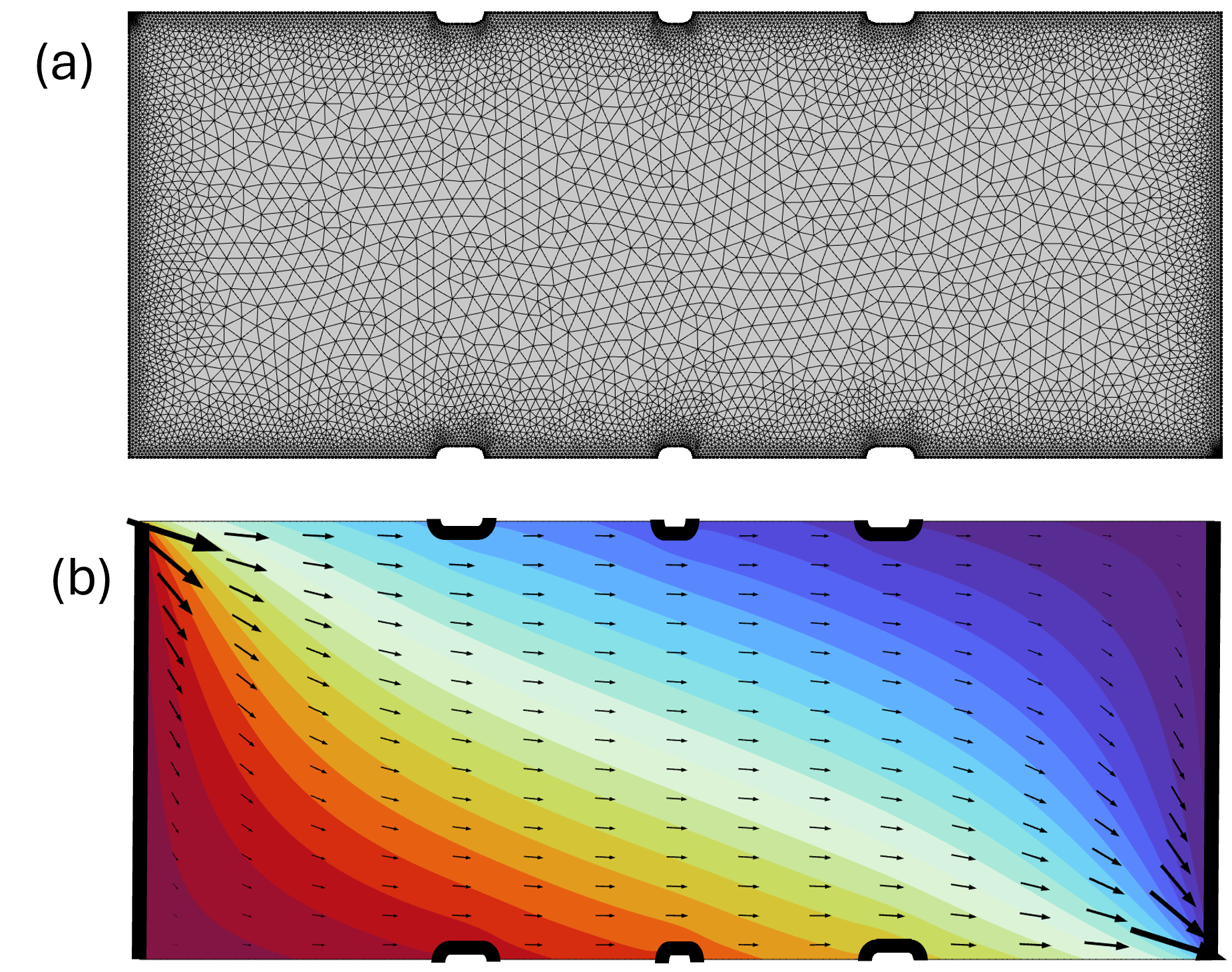}}%
	\caption{Results from the continuum Ohm's law model. (a) shows the mesh, where refinement
    near corners where current crowding occurs is important. In this case, this is around the upper left and lower right corners. (b) shows the electrostatic potential for $\sigma_{xy}/\sigma_{xx}=3.1$, along with arrows depicting the magnitude and direction of current. For this simulation, the source and drain are the left and right vertical edges, and all other edges' boundary condition is a vanishing normal component of current.}
	\label{fig:ohms}%
\end{figure}

We next present a description of the bulk dissipative model of transport.  We begin with a statement of linear response for charge transport (Ohm's law in continuum form):
\begin{eqnarray}
        {\bf J} &=&   \overleftrightarrow{\boldsymbol{\sigma}} {\bf E}  \label{eq:ohms}
\end{eqnarray}
We consider a two-dimensional system confined to the $xy$ plane. We assume the material is isotropic, so that $\sigma_{xx}=\sigma_{yy}$. The conductivity tensor includes anti-symmetric off-diagonal components, associated with the time-reversal symmetry breaking due to the out-of-plane magnetization. The conductivity tensor is then:
\begin{eqnarray}
    \overleftrightarrow{\boldsymbol{\sigma}}&=&  \begin{pmatrix}
    \sigma_{xx} & \sigma_{xy}\\
    -\sigma_{xy} & \sigma_{xx} \end{pmatrix}
\end{eqnarray}
Given specific boundary conditions, the electric field is uniquely determined by the continuity equation $\nabla \cdot {\bf J} = 0$, and the electric potential $V$ is in turn determined by ${\bf E} = -\nabla V$.  We utilize the finite-element method to evaluate Eq.~\ref{eq:ohms}. The contacts are located on the vertical edges, and on the notches along the horizontal edges (shown as dark lines on Fig.~\ref{fig:ohms}(b)). For larger values of $\sigma_{xy}/\sigma_{xx}$, we find mesh refinement near edges and charge current ``hot-spots'' is necessary. A representative mesh is shown in Fig.~\ref{fig:ohms}(a).  In this case, the hot-spots are at the upper left and lower right corners of the device (see Fig.~\ref{fig:ohms}(b)).

As described in Ref. \cite{rosen2022measured}, for $\sigma_{xy}/\sigma_{xx}\rightarrow \infty$, this model's predictions for transport experiments are indistinguishable from a conventional chiral edge state model.  Equivalently, both models yield an identical transmission or conductivity matrix in this limit.\\

\subsubsection{Continuum Ohm's law with non-chiral edge model}
To incorporate the continuum Ohm's law model with the non-chiral edge state model, we reformulate the transport description of the continuum Ohm's law in terms of an effective conductivity matrix. To do so, we perform a calculation with the boundary condition: $V_i\neq0,~V_j=0$ for $j\neq i$. The current collected by lead $j$ per voltage difference is denoted as a conductivity $G_{i,j}^{\rm bulk}$. To form the full $G_{i,j}^{\rm bulk}$ matrix, we perform calculations with eight different boundary conditions; each boundary condition is associated with a specific choice for current injection lead.


To combine this bulk transport with non-chiral edge state transport, we simply add the bulk conductivity matrix to the non-chiral model transmission matrix, as described in the previous subsection.  This description is clearly approximate, as we neglect coupling between bulk and non-chiral edge states. Generally, edge-bulk coupling is expected to increase conductivity, insofar as additional scattering channels provide more conduction paths. Therefore, we take our simplified treatment as a lower bound on the conductivity for a system with co-existing bulk and non-chiral edge states.

We account for the system geometry in our model of non-chiral edge states.  These states are subject to backscattering, and we therefore assume that the probability for inter-lead transmission is inversely proportional to the inter-lead distance.  This introduces a slight modification of the transmission probabilities; there are two different inter-lead distances in our system, with a ratio between them of 1.8.  Therefore, for the more closely-spaced leads $T_{i,j} = k$, while for the less closely-spaced leads, $T_{i,j}=k/1.8$.  This modification to match the device geometry makes only a slight impact on the results.

\section{results}

In this section, we present the results of the fitting for various models and discuss the measurement configuration for which the non-chiral state would be most easily identified.  Fig.~\ref{fig:data} compares the experimental data to the least-squares model fits, and Table~\ref{datatable} lists the corresponding fit parameters. The continuum Ohm’s-law model describes the data well across all measurement configurations, and the inclusion of a non-chiral edge state yields only marginal changes, suggesting the absence of non-chiral edge states in our Cr$_{0.12}$(Bi$_{0.26}$Sb$_{62}$)$_{2}$Te$_{3}$ samples. The chiral+non-chiral edge state only model does the poorest job of describing the experimental data. In the following subsection, we analyze the difference between these models in more detail. First, we discuss the chiral+non-chiral edge state model.

\begin{figure} [h!]
	{\includegraphics[scale=0.2,angle=0,trim={0.0cm 0.0cm 0.0cm 0.0cm},clip,width=.49\textwidth]{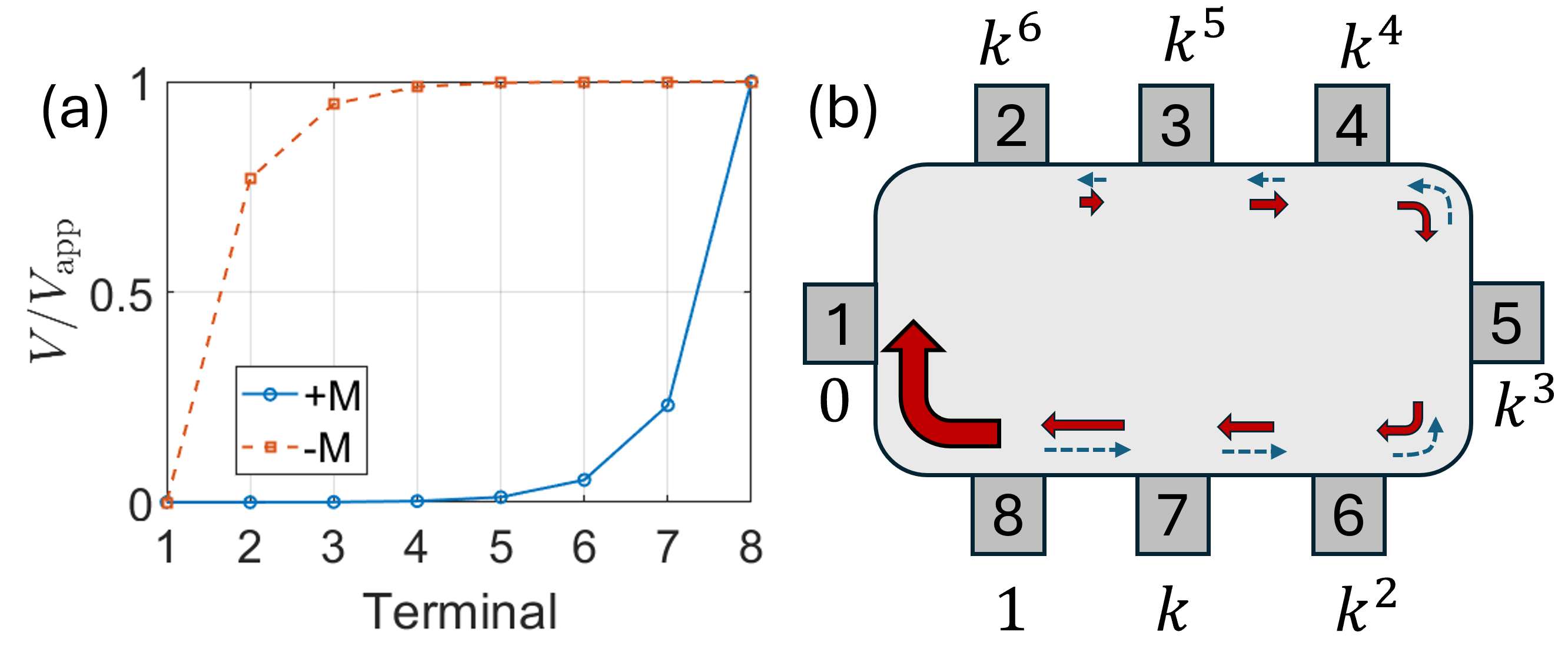}}%
	\caption{(a) Electrochemical potential versus terminal for both orientations of magnetization for the chiral+non-chiral edge state model.  The source and drain are terminals 8 and 1, respectively, and the non-chiral transmission probability is $k=0.3$. (b) depicts the terminal voltages (to lowest order in $k$, normalized by $V_{\rm app}$) and resulting current flow for the $+M$ configuration.  The bold red arrows depict chiral states, while the dashed blue arrows depict non-chiral states.}
	\label{fig:potential}
\end{figure}

For the chiral+non-chiral edge state model, the measurement configuration $(1,8;1,2)$ should provide the clearest signature of non-chiral edge state transport. Fig.~\ref{fig:potential}(a) shows the potentials obtained for the chiral+non-chiral edge state model with source and drain terminals of 1 and 8, for both $+M$ and $-M$.  For a perfectly quantized sample, the non-local resistance values are $R_{1,8;1,2}=0$ for $+M$ and $R_{1,8;1,2}=R_K$ for $-M$. For $+M$, the addition of a non-chiral state with transmission probability $k$ leads to $R_{1,8;1,2}=k^6R_K$ for $+M$ and $R_{1,8;1,2}=(1-k)R_K$ for $-M$.  The deviation from ideal quantized behavior is minimized at $+M$ (of order $k^6$), and maximized at $-M$ (of order $k$).  The signature of non-chiral edge states is therefore quite distinctive for this measurement configuration:  The discrepancy between the measured non-local resistance and the ideal, quantized value is very different for $+M$ and $-M$.  

To illustrate the origin of this behavior, we show a schematic of the lead potentials and current distribution for this measurement configuration in Fig. 5(b).  Leads 8 and 1 are source and drain, respectively, and the chiral edge state propagates clockwise.  In an ideal quantized case, the potential at lead 2 is zero.  The non-chiral edge state which propagates opposite to the chiral edge state modifies the lead 2 potential from its ideal, quantized value. For the case of $+M$, this corresponds to the counter-clockwise propagating non-chiral edge state.  Each intermediary lead between the source lead (lead 8) and lead 2 reduces the impact of the non-chiral edge state by a factor of $k$, leading to a contribution of $k^6$ on the potential of lead 2.  For $-M$, the clockwise non-chiral edge state between leads 1 and 2 affects the potential on lead 2.  There are no intermediary leads in this case, leading to a modification of the potential on lead 2 which is linear in $k$.  

In our measurements, the departures from the quantized value at $+M$ and $-M$ are $0.075~R_K$ and $-0.077~R_K$, respectively. In the model with both chiral and non-chiral edge states, the deviations are expected to scale as $\delta R_{1,8;1,2}(+M)\propto k^6 $ and $\delta R_{1,8;1,2}(-M)\propto -k$, as explained in the previous paragraph. The near equality of the observed offsets therefore provides immediate evidence that dissipation from non-chiral edge states is negligible.


\subsubsection{Statistical analysis of model fits}

We next quantify the difference in the model performance between the continuum Ohm's law model with and without the non-chiral edge state. The fitting obtained with the additional parameter describing the non-chiral edge state is slightly better (see the model errors in Table~\ref{datatable}.  This is not surprising: the inclusion of additional fitting parameters generally leads to a superior fit.  The key question is whether the improvement is significant enough to conclude that the additional parameter is warranted. To determine this, we perform additional statistical analysis.

We first review our least-squares model fitting procedure and parameter uncertainty determination.  Each model has a number of parameters which we arrange in a vector $\bf p$.  A given parameter choice yields a set of predictions for measured resistance values, $f_i({\bf p})$, where $i=1,...N$ labels the experiment number.  To find the optimum parameters, we perform a weighted least-square fit between measured values $y_i$ and model predictions $f_i$:
\begin{eqnarray}
    \sigma = \sqrt{ \frac{\sum_{i=1}^{N} w_i\left(y_i - f_i({\bf p})\right)^2 }{\sum_{i=1}^{N} w_i}}~,\label{eq:sigma}
\end{eqnarray}
where $w_i$ are the weights.  We set $w_i = 1/(\Delta R_i) ^2$, where $\Delta R_i$ is the experimental uncertainty of the $i^{\rm th}$ measurement.  The measurement uncertainty $\Delta R$ is determined through resistance measurements on calibrated resistors.  Using three different resistor values $(100,~ 1000,~ 25001)~\Omega$, we find that the measurement uncertainty obeys the empirical relation: $\Delta R  = 0.003\times R +R_0$, where $R_0=1.7~\Omega$.  There is additional uncertainty in the geometry used for the modeling, which is idealized and does not capture the exact geometry of the device.  To estimate the effect of non-ideal sample geometry, we compare the resistance values between nominally symmetry-equivalent measurements.  We find a discrepancy that is within the range of the $\Delta R$ specified above, and therefore use $\Delta R$ as our measurement of total uncertainty.


The best-fit parameters ${\bf p}_{\rm optimum}$ are chosen by minimizing the root mean square error.  To find the uncertainty of the least-square fit parameters,  we evaluate the Jacobian $J_{ij}=\delta f_i/\delta  p_j$ numerically at the optimum parameter ${\bf p}_{\rm optimum}$.  From this, we compute the covariance matrix of the model parameters, given by $C=\left(J^T J\right)^{-1}$.  The parameter uncertainty $\delta p_i$ is given by $\delta p_i =\sqrt{\sigma^2 C_{ii}}$, where $\sigma$ is evaluated at the optimum parameter.

Next, we describe the procedure for discriminating between different models' performance, focusing on the difference between the bulk continuum Ohm's law model with and without the non-chiral edge state.  In this case, the two-parameter bulk continuum Ohm's law model is a special case of the three-parameter bulk + non-chiral edge state model.  For such a case of ``nested models'', a commonly used approach is to utilize the $F$-statistic to determine if the additional parameters lead to a statistically significant improvement (see Sec. 6.1 of \cite{weisberg2005applied}). The $F$-statistic is given by:
\begin{eqnarray}
    F =\frac{ \left(\sigma^2_{A}-\sigma^2_{B}\right)/(n_A-n_B)}{\sigma^2 _B/(N-n_B)}
\end{eqnarray}
The subscript $A$ and $B$ label different models; in this case, model $A$ is a special case of model $B$.  $n_{A,B}$ is the number of parameters of model A,B ($n_B>n_A$), and $N$ is the number of fitted data points.  The $F$-statistic obtained from the data fits is compared to the $F$-distribution function $F_{n_A-n_B,N}$ to determine statistical significance.  For our case, $n_A-n_B=1$, $N=15$, leading to a requirement that $F>4.54$ for a $>95~\%$ confidence level in the statistical significance of the additional parameter.  For our data, we obtain $F=0.11$, which is very far from statistical significance.  Indeed, examination of the fits obtained with the two and three-parameter fit, shown in Fig. ~\ref{fig:non-chiral}b and c, shows that there is very marginal improvement with the addition of the non-chiral edge state to the model.  We therefore conclude that the data do not support the presence of a non-chiral edge state.


\begin{widetext}

\begin{figure}%
	{\includegraphics[scale=0.2,angle=0,trim={0.0cm 0.0cm 0.0cm 0.0cm},clip,width=1.05\textwidth]{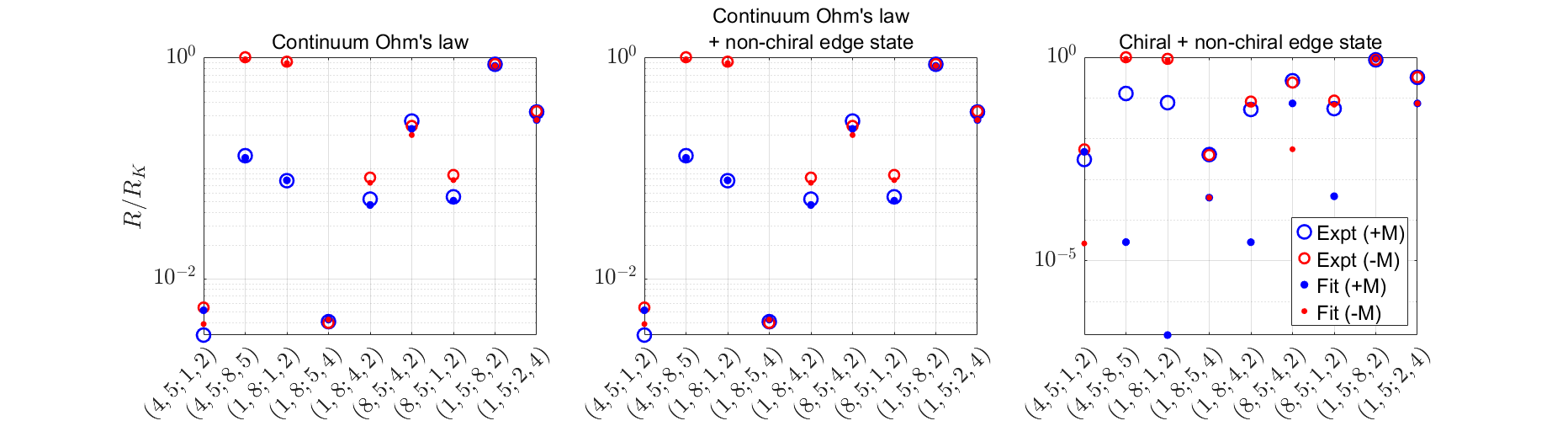}}%
	\caption{Results for the best fit of various models to the experimental data set.  }%
	\label{fig:data}%
\end{figure}


\begin{center}
\begin{table} 
\begin{tabular}{|c|| c |c| c|} 
\hline
   & Ohm's law & \makecell[c]{Continuum Ohm's law + \\ non-chiral edge state} & \makecell[c]{Chiral+non-chiral\\ edge state} \\ 
 \hline\hline
 \makecell[c]{Best fit \\ parameters}& 
 $\begin{aligned}
   ~~\tilde{\sigma}_{xx} &= 0.302 \pm 0.002~~ \\ 
   ~~\tilde{\sigma}_{xy} &= 0.937 \pm 0.005 
\end{aligned}$
  & $\begin{aligned}
   ~~~~\tilde{\sigma}_{xx} &= 0.302 \pm 0.003 ~~~~\\ 
   \tilde{\sigma}_{xy} &=  0.937 \pm 0.008 \\ 
   ~~k &=  (1.6 \pm 1.7) \times 10^{-3}~~
\end{aligned}$
 & $~~k= 0.141 \pm 0.011~~$ 
 \\ 
 \hline
 \makecell[c]{Model error \\ $\sigma$ [$\Omega$]} &  64.8 &  64.5 &  613.4  \\ 
 \hline
\end{tabular}
\caption{Fitting parameters and corresponding error for each model.}
\label{datatable}
\end{table}

\end{center}

\end{widetext}


\section{Discussion}

The results of the analysis show that, at 2 K, this material is well-described by a continuum Ohm's law, where all of the microscopic physics controlling transport behavior is effectively captured with two phenomenological parameters. This is consistent with the energy scales governing transport, where the activation energy is in the range of $1~{\rm K}$ \cite{fox2018part,okazaki2022quantum,pan2020probing}.  The activation energy is much smaller than the energy scale of the magnetic exchange gap, which is reflected in the much higher Curie temperature ($\approx 20~{\rm K}$) \cite{chang2013experimental,chang2015high}.  This mismatch of energy scales is well-known and has been quantitatively explored in previous works.  For example, scanning tunneling microscopy measurements used Landau level spectroscopy on ${\rm Cr_{0.08}(Bi_{0.1}Sb_{0.9})_{1.92}Te_3}$ to map the spatially inhomogeneous band edge and band gap \cite{chong2020severe}.  Statistical analysis revealed that the band gap between separated regions (that is, the difference between the conduction band edge energy at one position and the valence band edge energy at another position) is $100~\mu{\rm eV}$ (or $1~{\rm K}$) for regions within $<1~\mu{\rm m}$, potentially within the localization length of carriers \cite{fox2018part}.  The macroscopically-averaged transport of the device at $T=2~{\rm K}$ is then well-described by a bulk conduction model.  This transport picture has also been invoked to explain experiments on current-induced breakdown \cite{lippertz2022current}.  Previous non-local measurements at higher temperatures also find that bulk transport is largely responsible for the loss of quantization \cite{fijalkowski2021quantum}.  For these reasons, we believe that the current iteration of this materials class is unlikely to realize metrologically relevant levels of Hall quantization at 2 K.

In addition to characterizing the properties of a specific material, this work presents a systematic analysis of transport measurements in which the measurement configuration is the primary experimental parameter. We emphasize that this parameter can be particularly powerful for constraining transport models.  Prospective transport models should generally possess the same symmetry properties as the underlying system, so that discriminating between alternative models is essentially a {\it quantitative} exercise.  For this reason, larger data sets enable stronger conclusions. A consideration for future work is to utilize Hall bar geometries with lower symmetry, by, for example, intentionally displacing lead positions to remove spatial symmetries.  This increases the number of possible independent measurements and enables more data collection from a specific sample.

\section{Acknowledgements}

The sample was fabricated by Linsey K. Rodenbach and Molly P. Andersen in the Group of David Goldhaber-Gordon at Stanford from materials provided by Peng Zhang, Lixuan Tai, and Kang Wang at UCLA. We thank Alireza Panna and Michael Stewart for carefully reading the manuscript and providing insightful comments. FM acknowledges support under the Cooperative Research Agreement between the University of Maryland and the National Institute of Standards and Technology Physical Measurement Laboratory, Award 70NANB23H024, through the University of Maryland.

\section*{Data Availability}

The data presented in this work are not publicly available, but may be shared upon reasonable request. 

\appendix

\section{Symmetry Analysis}\label{app:symmetry}

In this appendix we describe the procedure for counting the number of independent measurements.  This is equal to the number of independent entries in the $T$ matrix.  To find this, a simple method is to average a general transmission matrix $T$ over the elements of the system's symmetry group $G$.  Begin with a general matrix $T_{\rm init}$ of an $N$-terminal device consisting of $N^2$ independent numbers.  Next, average over elements of the symmetry group:
\begin{eqnarray}
T_{\rm sym} = \frac{1}{N_G}\sum_{i\in G} U_i^{-1} T_{\rm init} U_i ,
\end{eqnarray}
where $N_G$ is the number of elements of the symmetry group. By construction, the resulting matrix $T_{\rm sym}$ satisfies all system symmetries.  The number of unique entries provides the number of symmetry-independent measurements.  For the symmetry group of the main text and an 8-terminal system, this procedure yields a matrix with 17 different entries.

	\bibliography{ref}{}

@book{weisberg2005applied,
  title={Applied linear regression},
  author={Weisberg, Sanford},
  volume={528},
  year={2005},
  publisher={John Wiley \& Sons}
}

@article{wang2013anomalous,
  title={Anomalous edge transport in the quantum anomalous Hall state},
  author={Wang, Jing and Lian, Biao and Zhang, Haijun and Zhang, Shou-Cheng},
  journal={Physical review letters},
  volume={111},
  number={8},
  pages={086803},
  year={2013},
  publisher={APS}
}

@article{kou2014scale,
  title={Scale-invariant quantum anomalous Hall effect in magnetic topological insulators beyond the two-dimensional limit},
  author={Kou, Xufeng and Guo, Shih-Ting and Fan, Yabin and Pan, Lei and Lang, Murong and Jiang, Ying and Shao, Qiming and Nie, Tianxiao and Murata, Koichi and Tang, Jianshi and others},
  journal={Physical review letters},
  volume={113},
  number={13},
  pages={137201},
  year={2014},
  publisher={APS}
}

@article{chang2015zero,
  title={Zero-field dissipationless chiral edge transport and the nature of dissipation in the quantum anomalous Hall state},
  author={Chang, Cui-Zu and Zhao, Weiwei and Kim, Duk Y and Wei, Peng and Jain, Jainendra K and Liu, Chaoxing and Chan, Moses HW and Moodera, Jagadeesh S},
  journal={Physical review letters},
  volume={115},
  number={5},
  pages={057206},
  year={2015},
  publisher={APS}
}

@article{wang2020demonstration,
  title={Demonstration of dissipative quasihelical edge transport in quantum anomalous hall insulators},
  author={Wang, Shu-Wei and Xiao, Di and Dou, Ziwei and Cao, Moda and Zhao, Yi-Fan and Samarth, Nitin and Chang, Cui-Zu and Connolly, Malcolm R and Smith, Charles G},
  journal={Physical review letters},
  volume={125},
  number={12},
  pages={126801},
  year={2020},
  publisher={APS}
}

@article{okazaki2022quantum,
  title={Quantum anomalous Hall effect with a permanent magnet defines a quantum resistance standard},
  author={Okazaki, Yuma and Oe, Takehiko and Kawamura, Minoru and Yoshimi, Ryutaro and Nakamura, Shuji and Takada, Shintaro and Mogi, Masataka and Takahashi, Kei S and Tsukazaki, Atsushi and Kawasaki, Masashi and others},
  journal={Nature Physics},
  volume={18},
  number={1},
  pages={25--29},
  year={2022},
  publisher={Nature Publishing Group UK London}
}

@article{patel2024zero,
  title={A zero external magnetic field quantum standard of resistance at the 10- 9 level},
  author={Patel, DK and Fijalkowski, KM and Kruskopf, M and Liu, N and G{\"o}tz, M and Pesel, E and Jaime, M and Klement, M and Schreyeck, S and Brunner, K and others},
  journal={Nature Electronics},
  volume={7},
  number={12},
  pages={1111--1116},
  year={2024},
  publisher={Nature Publishing Group UK London}
}

@article{rodenbach2022metrological,
  title={Metrological assessment of quantum anomalous Hall properties},
  author={Rodenbach, Linsey K and Panna, Alireza R and Payagala, Shamith U and Rosen, Ilan T and Andersen, Molly P and Zhang, Peng and Tai, Lixuan and Wang, Kang L and Jarrett, Dean G and Elmquist, Randolph E and others},
  journal={Physical Review Applied},
  volume={18},
  number={3},
  pages={034008},
  year={2022},
  publisher={APS}
}

@article{rodenbach2025unified,
  title={A unified realization of electrical quantities from the quantum International System of Units},
  author={Rodenbach, Linsey K and Underwood, Jason M and Tran, Ngoc Thanh Mai and Panna, Alireza R and Andersen, Molly P and Barcikowski, Zachary S and Payagala, Shamith U and Zhang, Peng and Tai, Lixuan and Wang, Kang L and others},
  journal={Nature Electronics},
  pages={1--9},
  year={2025},
  publisher={Nature Publishing Group UK London}
}

@article{callegaro2025quahmet,
  title={QuAHMET: Quantum anomalous Hall effect materials and devices for metrology},
  author={Callegaro, Luca and Marzano, Martina and Medved, Juan and Gould, Charles and Hoffmann, Johannes and Huang, Nathaniel and Kaneko, Nobu-Hisa and Kucera, Jan and Molenkamp, Laurens W and Onbasli, Mehmet Cengiz and others},
  journal={Measurement: Sensors},
  volume={38},
  pages={101437},
  year={2025},
  publisher={Elsevier}
}

@article{fox2018part,
  title={Part-per-million quantization and current-induced breakdown of the quantum anomalous Hall effect},
  author={Fox, Eli J and Rosen, Ilan T and Yang, Yanfei and Jones, George R and Elmquist, Randolph E and Kou, Xufeng and Pan, Lei and Wang, Kang L and Goldhaber-Gordon, D},
  journal={Physical Review B},
  volume={98},
  number={7},
  pages={075145},
  year={2018},
  publisher={APS}
}

@article{huang2025quantum,
  title={Quantum anomalous Hall effect for metrology},
  author={Hu{\'a}ng, Nathaniel J and Boland, Jessica L and Fijalkowski, Kajetan M and Gould, Charles and Hesjedal, Thorsten and Kazakova, Olga and Kumar, Susmit and Scherer, Hansj{\"o}rg},
  journal={Applied Physics Letters},
  volume={126},
  number={4},
  year={2025},
  publisher={AIP Publishing}
}

@article{chang2023colloquium,
  title={Colloquium: Quantum anomalous hall effect},
  author={Chang, Cui-Zu and Liu, Chao-Xing and MacDonald, Allan H},
  journal={Reviews of Modern Physics},
  volume={95},
  number={1},
  pages={011002},
  year={2023},
  publisher={APS}
}

@article{zhang2010crossover,
  title={Crossover of the three-dimensional topological insulator Bi2Se3 to the two-dimensional limit},
  author={Zhang, Yi and He, Ke and Chang, Cui-Zu and Song, Can-Li and Wang, Li-Li and Chen, Xi and Jia, Jin-Feng and Fang, Zhong and Dai, Xi and Shan, Wen-Yu and others},
  journal={Nature Physics},
  volume={6},
  number={8},
  pages={584--588},
  year={2010},
  publisher={Nature Publishing Group UK London}
}

@article{rodenbach2021bulk,
  title={Bulk dissipation in the quantum anomalous Hall effect},
  author={Rodenbach, Linsey K and Rosen, Ilan T and Fox, Eli J and Zhang, Peng and Pan, Lei and Wang, Kang L and Kastner, Marc A and Goldhaber-Gordon, David},
  journal={APL Materials},
  volume={9},
  number={8},
  year={2021},
  publisher={AIP Publishing}
}

@article{rosen2022measured,
  title={Measured potential profile in a quantum anomalous Hall system suggests bulk-dominated current flow},
  author={Rosen, Ilan T and Andersen, Molly P and Rodenbach, Linsey K and Tai, Lixuan and Zhang, Peng and Wang, Kang L and Kastner, MA and Goldhaber-Gordon, David},
  journal={Physical review letters},
  volume={129},
  number={24},
  pages={246602},
  year={2022},
  publisher={APS}
}

@misc{footnote,
note={Note that the temperature-dependent longitudinal conductivity often exhibits a crossover between variable-range-hopping at lowest temperatures to thermally-activated transport at moderate temperatures.}
}

@article{pan2020probing,
  title={Probing the low-temperature limit of the quantum anomalous Hall effect},
  author={Pan, Lei and Liu, Xiaoyang and He, Qing Lin and Stern, Alexander and Yin, Gen and Che, Xiaoyu and Shao, Qiming and Zhang, Peng and Deng, Peng and Yang, Chao-Yao and others},
  journal={Science advances},
  volume={6},
  number={25},
  pages={eaaz3595},
  year={2020},
  publisher={American Association for the Advancement of Science}
}

@article{ferguson2023direct,
  title={Direct visualization of electronic transport in a quantum anomalous Hall insulator},
  author={Ferguson, GM and Xiao, Run and Richardella, Anthony R and Low, David and Samarth, Nitin and Nowack, Katja C},
  journal={Nature Materials},
  volume={22},
  number={9},
  pages={1100--1105},
  year={2023},
  publisher={Nature Publishing Group UK London}
}

@article{lippertz2022current,
  title={Current-induced breakdown of the quantum anomalous Hall effect},
  author={Lippertz, Gertjan and Bliesener, Andrea and Uday, Anjana and Pereira, Lino MC and Taskin, AA and Ando, Yoichi},
  journal={Physical Review B},
  volume={106},
  number={4},
  pages={045419},
  year={2022},
  publisher={APS}
}

@article{fijalkowski2021quantum,
  title={Quantum anomalous Hall edge channels survive up to the Curie temperature},
  author={Fijalkowski, Kajetan M and Liu, Nan and Mandal, Pankaj and Schreyeck, Steffen and Brunner, Karl and Gould, Charles and Molenkamp, Laurens W},
  journal={Nature communications},
  volume={12},
  number={1},
  pages={5599},
  year={2021},
  publisher={Nature Publishing Group UK London}
}

@article{chang2013experimental,
  title={Experimental observation of the quantum anomalous Hall effect in a magnetic topological insulator},
  author={Chang, Cui-Zu and Zhang, Jinsong and Feng, Xiao and Shen, Jie and Zhang, Zuocheng and Guo, Minghua and Li, Kang and Ou, Yunbo and Wei, Pang and Wang, Li-Li and others},
  journal={Science},
  volume={340},
  number={6129},
  pages={167--170},
  year={2013},
  publisher={American Association for the Advancement of Science}
}

@article{chang2015high,
  title={High-precision realization of robust quantum anomalous Hall state in a hard ferromagnetic topological insulator},
  author={Chang, Cui-Zu and Zhao, Weiwei and Kim, Duk Y and Zhang, Haijun and Assaf, Badih A and Heiman, Don and Zhang, Shou-Cheng and Liu, Chaoxing and Chan, Moses HW and Moodera, Jagadeesh S},
  journal={Nature materials},
  volume={14},
  number={5},
  pages={473--477},
  year={2015},
  publisher={Nature Publishing Group UK London}
}

@article{chong2020severe,
  title={Severe dirac mass gap suppression in sb2te3-based quantum anomalous hall materials},
  author={Chong, Yi Xue and Liu, Xiaolong and Sharma, Rahul and Kostin, Andrey and Gu, Genda and Fujita, K and Davis, JC S{\'e}amus and Sprau, Peter O},
  journal={Nano Letters},
  volume={20},
  number={11},
  pages={8001--8007},
  year={2020},
  publisher={ACS Publications}
}

@article{buttiker1986four,
  title={Four-terminal phase-coherent conductance},
  author={B{\"u}ttiker, M},
  journal={Physical review letters},
  volume={57},
  number={14},
  pages={1761},
  year={1986},
  publisher={APS}
}

@article{zhao20243d,
  title={3D Quantum Anomalous Hall Effect in Magnetic Topological Insulator Trilayers of Hundred-Nanometer Thickness},
  author={Zhao, Yi-Fan and Zhang, Ruoxi and Sun, Zi-Ting and Zhou, Ling-Jie and Zhuo, Deyi and Yan, Zi-Jie and Yi, Hemian and Wang, Ke and Chan, Moses HW and Liu, Chao-Xing and others},
  journal={Advanced Materials},
  volume={36},
  number={13},
  pages={2310249},
  year={2024},
  publisher={Wiley Online Library}
}

\end{document}